\documentclass[11pt]{article}%
\usepackage{amsmath}
\usepackage{hyperref}
\usepackage{amsfonts}
\usepackage{amssymb}
\usepackage{graphicx}
\usepackage{fullpage}
\usepackage{pgf}
\usepackage{tikz}
\usepackage{xy}
\usepackage{ifpdf}
\usepackage{pdfpages}
\usepackage{siunitx}%
\providecommand{\U}[1]{\protect\rule{.1in}{.1in}}
\usetikzlibrary{calc,positioning,fit}
\usetikzlibrary{fit,svg.path,positioning}
\newtheorem{theorem}{Theorem}

\newtheorem{problem}[theorem]{Problem}

\xyoption{matrix}
\xyoption{frame}
\xyoption{arrow}
\xyoption{arc}
\ifpdf
\else
\PackageWarningNoLine{Qcircuit}{Qcircuit is loading in Postscript mode.  The Xy-pic options ps and dvips will be loaded.  If you wish to use other Postscript drivers for Xy-pic, you must modify the code in Qcircuit.tex}
\xyoption{ps}
\xyoption{dvips}
\fi
\entrymodifiers={!C\entrybox}

\begin{document}

\title{Open Problems Related to Quantum\ Query Complexity}
\author{Scott Aaronson\thanks{University of Texas at Austin. \ Email:
aaronson@utexas.edu. \ Supported by a Vannevar Bush Fellowship from the US
Department of Defense, a Simons Investigator Award, and the Simons
\textquotedblleft It from Qubit\textquotedblright\ collaboration.}}
\date{}
\maketitle

\begin{abstract}
I offer a case that quantum query complexity still has loads of enticing and
fundamental open problems---from relativized $\mathsf{QMA}$ versus
$\mathsf{QCMA}$ and $\mathsf{BQP}$ versus $\mathsf{IP}$, to time/space
tradeoffs for collision and element distinctness, to polynomial degree versus
quantum query complexity for partial functions, to the Unitary Synthesis
Problem and more.

\end{abstract}

\section{Introduction}

Quantum query complexity (see \cite{bw} for a classic survey) is the study of
how many queries a quantum computer needs to make to an input string $X$ to
learn various properties of $X$. \ The key here is that a single query can
access multiple bits of $X$, one in each branch of a superposition state.
\ For over thirty years, this subject has been a central source of what we
know about both the capabilities and the limitations of quantum computers.

In my view, there are two reasons why query complexity has played such an
important role in quantum computing theory as a whole. \ First, it so happens
that most of the famous quantum algorithms---including Deutsch-Jozsa
\cite{dj}, Bernstein-Vazirani \cite{bv}, Simon \cite{simon}, Shor \cite{shor},
and Grover \cite{grover}---fit naturally into the query complexity framework,
or (in the case of Shor's algorithm) have a central component that does.
\ Second, query complexity lets us prove not only upper bounds, but also
nontrivial and informative lower bounds---as illustrated by the seminal 1994
theorem of Bennett, Bernstein, and Vazirani \cite{bbbv} that a quantum
computer needs $\Omega(\sqrt{N})$ queries to search an unordered list of size
$N$ for a single \textquotedblleft marked item.\textquotedblright\ \ This both
demonstrated the optimality of Grover's algorithm, two years before that
algorithm had been discovered to exist (!), and showed the existence of an
oracle relative to which $\mathsf{NP}\not \subset \mathsf{BQP}$.

Of course, oracle separations sometimes mislead us about the \textquotedblleft
real world,\textquotedblright\ where no oracles are present---a famous example
being the 1990 $\mathsf{IP}=\mathsf{PSPACE}$ theorem \cite{shamir}. \ Even
after a quarter century, though, non-relativizing techniques (i.e., techniques
that transcend query complexity) have made only minor inroads into quantum
complexity theory, at least outside the usual place where those techniques
have shined: namely, the study of interactive proof systems.

Yet today, some in our field seem to have the impression that quantum query
complexity is more-or-less a closed subject. \ Certainly, some of quantum
computing theorists' attention has understandably shifted to other topics,
from the theoretical foundations of quantum supremacy experiments
\cite{lbr:supremacy}, to potential near-term or \textquotedblleft
NISQ\textquotedblright\ quantum algorithms \cite{preskill:nisq}, to the quest
to prove a quantum PCP Theorem \cite{aav:qpcp}. \ And certainly, many of the
great open problems of quantum query complexity from circa 2000 \textit{were}
ultimately solved: to give some well-known examples, the quantum query
complexities of the collision and element distinctness problems \cite{as}\ and
of evaluating read-once formulas \cite{reichardt:reflections}; the optimal
separation between deterministic and quantum query complexities of total
Boolean functions \cite{abblss,abkrt}; and an oracle separation between
$\mathsf{BQP}$ and the polynomial hierarchy \cite{raztal}.

Nevertheless, in this article I'd like to make the case that open problems
abound in quantum query complexity---and I don't mean detail problems, of
tightening some bound to remove a logarithmic factor, but big, juicy,
important problems. \ Some of my problems are old and well-known; others are
obscure; still others, as far as I know, appear here in writing for the first time.

\section{$\mathsf{QMA}$ and Oracle Separations\label{QMA}}

Recall that $\mathsf{QMA}$, Quantum Merlin Arthur, is the class of languages
$L$\ for which membership in $L$ can be proven via a polynomial-size quantum
witness state $\left\vert \varphi\right\rangle $\ that's verified in quantum
polynomial time. \ In 2002, Aharonov and Naveh \cite{an}\ defined
$\mathsf{QCMA}$, or Quantum Classical Merlin Arthur, to be the subclass of
$\mathsf{QMA}$\ where the witness $\left\vert \varphi\right\rangle =\left\vert
w\right\rangle $\ is required to be a classical basis state (i.e., a string).
\ Ever since, one of the fundamental problems of quantum complexity theory has
been whether $\mathsf{QMA}=\mathsf{QCMA}$. \ One distinctive feature of this
question is that even its query complexity analogue remains open:

\begin{problem}
Is there an oracle relative to which $\mathsf{QMA}\neq\mathsf{QCMA}$?
\end{problem}

In 2007, Greg Kuperberg and I \cite{ak}\ at least showed that there's a
\textit{quantum} oracle $U$---that is, a collection of unitary transformations
provided as black boxes---such that $\mathsf{QMA}^{\mathcal{U}}\neq
\mathsf{QCMA}^{\mathcal{U}}$. \ This was the first use of quantum oracles in
complexity theory; their use has since become standard. \ But the question of
whether $\mathsf{QMA}$ and $\mathsf{QCMA}$\ can be separated by a
\textquotedblleft standard\textquotedblright\ oracle remained wide open. \ In
2015, Fefferman and Kimmel \cite{feffermankimmel} showed that there's a
\textquotedblleft randomized, in-place\textquotedblright\ classical oracle
relative to which $\mathsf{QMA}\neq\mathsf{QCMA}$, but for proving a
conventional classical oracle separation, I believe the best candidate we have
remains the \textquotedblleft component mixers problem\textquotedblright%
\ introduced in 2011 by Lutomirski \cite{lutomirski:scp}.

Here is another fundamental problem that's remained open about $\mathsf{QMA}%
$\ and oracle separations:

\begin{problem}
Is there an oracle---even a quantum oracle---relative to which $\mathsf{QMA}%
\neq\mathsf{QMA}\left(  2\right)  $?
\end{problem}

Here, $\mathsf{QMA}\left(  2\right)  $\ is the analogue of $\mathsf{QMA}$\ to
allow \textit{two} unentangled Merlins, so that Arthur can always assume that
the witness state he receives is a tensor product $\left\vert \psi
\right\rangle \otimes\left\vert \varphi\right\rangle $ across two
polynomial-size registers. \ Despite 18 years of work on this class, the only
inclusions known are still the obvious ones, $\mathsf{QMA}\subseteq
\mathsf{QMA}\left(  2\right)  \subseteq\mathsf{NEXP}$. \ Furthermore, unlike
with the $\mathsf{QMA}$ vs. $\mathsf{QCMA}$\ problem, here we do not even have
a \textit{quantum} oracle relative to which $\mathsf{QMA}\neq\mathsf{QMA}%
\left(  2\right)  $. \ Watrous (see \cite{abdfs}) conjectured that there is no
quantum channel that takes polynomially many qubits as input, produces
polynomially many qubits as output, always produces an approximately separable
state on two registers as its output, and can approximately produce
\textit{any} separable state. \ Proving Watrous's \textquotedblleft no
disentanglers conjecture\textquotedblright\ is a prerequisite to separating
$\mathsf{QMA}$ from $\mathsf{QMA}\left(  2\right)  $\ query complexity, since
were his conjecture false, we could always use a $\mathsf{QMA}$\ witness to
simulate a $\mathsf{QMA}\left(  2\right)  $\ witness. \ See Harrow, Natarajan,
and Wu \cite{hnw}\ for the best current progress toward proving Watrous's conjecture.

\section{Query/Space Tradeoffs}

In the \textit{collision} problem, we're given black-box access to a function
$f:\left\{  1,\ldots,n\right\}  \rightarrow\left\{  1,\ldots,m\right\}  $
(where $n$ is even and $m\geq n$), and are asked to decide whether $f$ is
1-to-1 or 2-to-1, promised that one of those is the case. \ In the
\textit{element distinctness} problem, we're given black-box access to a
function $f:\left\{  1,\ldots,n\right\}  \rightarrow\left\{  1,\ldots
,m\right\}  $, with no promise, and are simply asked whether $f$ is 1-to-1.

\begin{problem}
What are the optimal tradeoffs between the number of queries used by a quantum
algorithm to solve the collision or the element distinctness problems, and the
number of qubits or classical bits of memory?
\end{problem}

Brassard, H\o yer, Tapp \cite{bht} gave a quantum algorithm for the collision
problem that uses $O\left(  n^{1/3}\right)  $\ quantum queries, as well as
$O\left(  n^{1/3}\right)  $ bits of classical memory and $O\left(  \log
n\right)  $\ qubits. \ Six years later, Ambainis \cite{ambainis:walk}\ gave a
quantum algorithm for element distinctness that uses $O\left(  n^{2/3}\right)
$\ quantum queries and $O\left(  n^{2/3}\right)  $ qubits. \ The 2002
collision lower bound by me and Yaoyun Shi \cite{as} showed that both of these
algorithms were optimal in terms of queries, thereby settling the problems'
quantum query complexity.

Here, though, we're asking whether a quantum algorithm for these problems
could achieve near-optimal query complexity while also using a small
\textit{memory}. \ Note that, by using Grover's algorithm, we could solve the
collision problem using only $O\left(  \log n\right)  $\ qubits in total, but
then we'd need $O(\sqrt{n})$\ queries\ rather than $O\left(  n^{1/3}\right)
$. \ For element distinctness, even supposing that we need a large memory, it
would also be interesting to know whether the memory needs to be made of
qubits, or whether coherently-queryable classical bits (a so-called
\textquotedblleft qRAM\textquotedblright) would suffice.

At present, unfortunately, the only techniques that we have for proving
quantum lower bounds that trade off space with query complexity, seem to apply
only to problems with many bits of output, such as sorting a list \cite{ksw}.
\ Proving such lower bounds for decision problems, like collision or element
distinctness, will probably require the invention of new techniques.

\section{Maximal Separations}

Given a total Boolean function $f:\left\{  0,1\right\}  ^{n}\rightarrow
\left\{  0,1\right\}  $, we denote by $\operatorname*{D}\left(  f\right)  $,
$\operatorname*{R}\left(  f\right)  $, and $\operatorname*{Q}\left(  f\right)
$\ the deterministic, (bounded-error) randomized, and (bounded-error) quantum
query complexities of $f$ respectively. \ In their seminal 1998 paper, Beals
et al.\ \cite{bbcmw} showed that $\operatorname*{D}\left(  f\right)  =O\left(
\operatorname*{Q}\left(  f\right)  ^{6}\right)  $ for all $f$. \ This stood as
the best known relationship between $\operatorname*{D}\left(  f\right)  $\ and
$\operatorname*{Q}\left(  f\right)  $ until extremely recently, when, building
on Huang's breakthrough proof of the Sensitivity Conjecture \cite{huang}, some
of us \cite{abkrt}\ showed that $\operatorname*{D}\left(  f\right)  =O\left(
\operatorname*{Q}\left(  f\right)  ^{4}\right)  $ for all total Boolean
functions $f$.

In the other direction, until 2015 it was widely believed that the largest
possible gap between classical and quantum query complexities for total
Boolean functions was quadratic, and was achieved by Grover's algorithm
applied to the $n$-bit OR function. \ But Ambainis et al.\ \cite{abblss} then
refuted that conjecture, by giving an example of a Boolean function $f$ for
which $\operatorname*{D}\left(  f\right)  \approx\operatorname*{Q}\left(
f\right)  ^{4}$. \ Not long afterward, Ben-David \cite{bendavid}\ (see also
Aaronson, Ben-David, and Kothari \cite{abk}) gave an example of an $f$ for
which $\operatorname*{R}\left(  f\right)  \approx\operatorname*{Q}\left(
f\right)  ^{2.5}$, thereby showing that Ambainis et al.'s separation was not
just an artifact of ignoring classical randomized algorithms. \ Ben-David's
result was recently improved to give functions $f$\ for which
$\operatorname*{R}\left(  f\right)  \approx\operatorname*{Q}\left(  f\right)
^{8/3}$\ \cite{tal:towards}\ and even $\operatorname*{R}\left(  f\right)
\approx\operatorname*{Q}\left(  f\right)  ^{3}$ \cite{bansalsinha,sherstovsw}.

Yet all this progress, as dramatic as it's been, still leaves a gap between
$3$ and $4$ in the exponent of the optimal separation between
$\operatorname*{R}\left(  f\right)  $\ and $\operatorname*{Q}\left(  f\right)
$.

\begin{problem}
What is the largest possible gap between $\operatorname*{R}\left(  f\right)
$\ and $\operatorname*{Q}\left(  f\right)  $, for a total Boolean function $f$?
\end{problem}

\section{Degree of Partial Functions}

Let $f:S\rightarrow\left\{  0,1\right\}  $ be a partial Boolean function,
where $S\subseteq\left\{  0,1\right\}  ^{n}$. \ Define the \textit{approximate
degree} of $f$, or $\widetilde{\deg}\left(  f\right)  $, to be the minimum
degree of a real polynomial $p:\mathbb{R}^{n}\rightarrow\mathbb{R}$ such that

\begin{enumerate}
\item[(i)] $\left\vert p\left(  x\right)  -f\left(  x\right)  \right\vert
\leq\frac{1}{3}$ for all $x\in S$, and

\item[(ii)] $p\left(  x\right)  \in\left[  0,1\right]  $ for all $x\in\left\{
0,1\right\}  ^{n}$.
\end{enumerate}

The seminal 1998 work of Beals at al. \cite{bbcmw}\ showed that
$\widetilde{\deg}\left(  f\right)  \leq2\operatorname*{Q}\left(  f\right)
$\ for all $f$, where $\operatorname*{Q}\left(  f\right)  $ is bounded-error
quantum query complexity. \ This is simply because the acceptance probability
of a $T$-query quantum algorithm \textit{is} a real polynomial of degree at
most $2T$. \ Beals et al.'s result was the beginning of the wildly-successful
\textit{polynomial method} in quantum complexity theory, whose central idea is
that to lower-bound quantum query complexity, it suffices to lower-bound
approximate degree.

In 2003, Ambainis \cite{ambainis:deg} showed that there can be small
polynomial gaps between $\widetilde{\deg}\left(  f\right)  $\ and
$\operatorname*{Q}\left(  f\right)  $\ for total Boolean functions $f$---and
thus, \textquotedblleft the polynomial method is not tight.\textquotedblright%
\ \ Ben-David, Kothari, and I \cite{abk}\ later improved this to get a
fourth-power gap between $\widetilde{\deg}\left(  f\right)  $\ and
$\operatorname*{Q}\left(  f\right)  $, which is tight by the recent work of
Ben-David, Kothari, Rao, Tal, and me \cite{abkrt}.

I ask about the situation for \textit{partial} Boolean functions:

\begin{problem}
What is the largest possible gap between $\operatorname*{Q}\left(  f\right)  $
and $\widetilde{\deg}\left(  f\right)  $ for partial $f$? \ Can the gap even
be exponential?
\end{problem}

Note that, if it weren't for requirement (ii)---namely, that the polynomial
must be bounded in $\left[  0,1\right]  $\ even on inputs that violate the
promise---we'd have a degree-$1$ polynomial representing the $n$-bit OR
function, whose quantum query complexity is $\Theta(\sqrt{n})$.

\section{Unitary Synthesis Problem}

In 2007, Greg Kuperberg and I \cite{ak}\ raised the following question:

\begin{problem}
For every $n$-qubit unitary transformation $U$, does there exist an oracle
$A:\left\{  0,1\right\}  ^{\ast}\rightarrow\left\{  0,1\right\}  $\ such that
a $\mathsf{BQP}^{A}$\ machine can implement $U$?\footnote{Technically,
Kuperberg and I only asked whether this is true for a Haar-random U, but I
expect the Haar-random case to be essentially the hardest case.}
\end{problem}

In my 2016 Barbados lecture notes \cite{aarbados}, I took to calling this the
\textquotedblleft Unitary Synthesis Problem.\textquotedblright\ \ It remains
wide open.

For comparison, it's not hard to show that, for every $n$-qubit state
$\left\vert \psi\right\rangle $, there exists an oracle $A$ such that a
$\mathsf{BQP}^{A}$ machine can prepare $\left\vert \psi\right\rangle $.
\ Indeed, for every $n$-qubit unitary $U$ and every polynomial $p$, there
exists an oracle $A$ such that a $\mathsf{BQP}^{A}$ machine can simulate the
behavior of $U$ on any chosen $p\left(  n\right)  $ basis states. However,
extending this construction to simulate $U$ on \textit{all} states seems to
entail exponentially many queries to $A$.

While it might sound esoteric, the Unitary Synthesis Problem has turned up
again and again---for example, in the study of the nonabelian hidden subgroup
problem \cite{ehk}, of decoding Hawking radiation from a black hole
\cite{harlowhayden,aarbados}, and of schemes for quantum copy-protection and
quantum money \cite{aar:qcopy,aarbados}. \ In each of those topics, one is
interested in certain complicated $n$-qubit unitary transformations $U$---and
especially, whether or not those $U$'s have polynomial-size quantum circuits.
\ The question arises: could we at least show that small quantum circuits
would exist \textit{if} (say) $\mathsf{P}=\mathsf{PSPACE}$, or some other
classical complexity classes dramatically collapsed? \ While the implication
isn't immediate, a positive answer to the Unitary Synthesis Problem would
strongly suggest that the answer was yes. For what it's worth, though, my
conjecture is that the answer is negative---in which case, the study of
quantum circuit complexity \textit{cannot} be so easily related to classical
complexity theory.

\section{Verifiability of Quantum Computing}

Let $\mathsf{IP}$\ be the class of languages that admit classical interactive
proofs. \ In the unrelativized world, $\mathsf{IP}=\mathsf{PSPACE}%
$\ \cite{shamir}, but it's well-known that the situation relative to oracles
can be dramatically different \cite{fs}. \ Thus we ask:

\begin{problem}
Does there exist an oracle $A$ such that $\mathsf{BQP}^{A}\not \subset
\mathsf{IP}^{A}$?
\end{problem}

In my view, the Forrelation problem, which I \cite{aar:ph}\ introduced in
2009, and which Raz and Tal \cite{raztal} used in 2018 to give an oracle
relative to which $\mathsf{BQP}\not \subset \mathsf{PH}$, provides a
compelling candidate for an oracle relative to which $\mathsf{BQP}%
\not \subset \mathsf{IP}$ as well. \ However, showing that Forrelation is not
in $\mathsf{IP}$ will require a new circuit lower bound---one that talks about
circuits with \textquotedblleft expectation\textquotedblright\ and
\textquotedblleft maximization\textquotedblright\ gates, rather than
$\mathsf{AC}^{0}$ circuits with AND, OR, and NOT gates. \ As far as I know,
Aiello, Goldwasser, and H\aa stad \cite{agh} proved what's still the best
known lower bound against expectation/maximization circuits in 1989, when they
gave an oracle relative to which more rounds give interactive protocols more power.

Now let $\mathsf{IP}_{\mathsf{BQP}}$ be the subclass of $\mathsf{BQP}$
consisting of all languages for which a $\mathsf{BQP}$ prover can convince a
$\mathsf{BPP}$ verifier of a \textquotedblleft yes\textquotedblright\ answer,
through polynomially many rounds of classical interaction.

\begin{problem}
Is there at least an oracle relative to which $\mathsf{BQP}\neq\mathsf{IP}%
_{\mathsf{BQP}}$?
\end{problem}

Besides Forrelation, even the complement of Simon's Problem (i.e., output
\textquotedblleft yes\textquotedblright\ if $f$ is a 1-to-1 function, or
\textquotedblleft no\textquotedblright\ if $f$ satisfies the Simon promise,
promised that one of these is the case) seems like a good candidate for an
oracle problem in $\mathsf{BQP}$ but not in $\mathsf{IP}_{\mathsf{BQP}}$. \ In
the Simon example, note that there \textit{is} an interactive protocol, based
on $\mathsf{AM}$\ approximate counting---it just doesn't seem to be a protocol
for which a $\mathsf{BQP}$ machine could implement the prover's strategy.

Finally, a question that was implicit in our previous one:

\begin{problem}
Are the celebrated protocols for blind and verified quantum computation, due
to Broadbent et al.\ \cite{bfk}, Aharonov et al.\ \cite{abem}, and Mahadev
\cite{mahadev}, inherently non-relativizing?
\end{problem}

Certainly these protocols don't \textit{manifestly} work relative to arbitrary
oracles, but are there variants of the protocols that do?

\section{Glued Trees}

In 2002, Childs et al.\ \cite{ccdfgs} gave a celebrated quantum walk algorithm
that, informally, gets from the leftmost to the rightmost vertex in the
following $\exp\left(  n\right)  $-sized graph, in only $\operatorname*{poly}%
\left(  n\right)  $ time and with $\frac{1}{\operatorname*{poly}\left(
n\right)  }$\ success probability:

\includegraphics{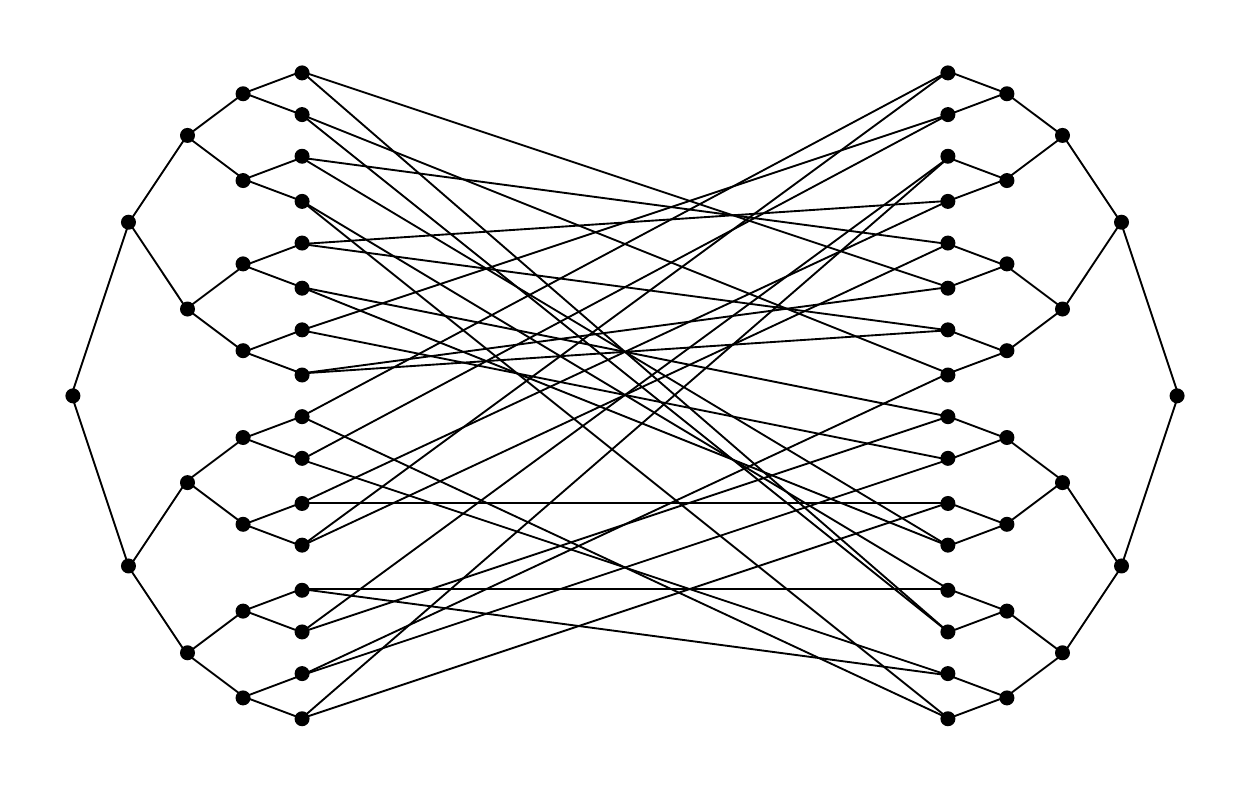}

By contrast, they showed that a randomized algorithm needs $2^{\Omega\left(
n\right)  }$ queries to an oracle encoding the graph to solve the same problem
(improved by Fenner and Zhang \cite{fennerzhang}\ to $\widetilde{\Omega
}\left(  2^{n/2}\right)  $).

\begin{problem}
Suppose, however, that we actually want to \textit{find a path} from left to
right. \ Does even a quantum computer need $2^{\Omega\left(  n\right)  }$
queries for that task?
\end{problem}

Certainly, if we try to measure the state of the quantum walk algorithm to
reveal such a path, we'll destroy the quantum interference that causes that
algorithm to succeed. \ But this, of course, doesn't show that no other
quantum algorithm is possible. \ It seems to me that a lower bound---showing
that a quantum algorithm can't efficiently find even a single left-right path,
even though it can traverse exponentially many such paths in
superposition---would be a striking algorithmic version of wave/particle duality.

\section{Comparing Query Models}

Given a function $f$, the usual model of quantum query complexity is that we
get access to an oracle that maps basis states of the form $\left\vert
x,a\right\rangle $ to basis states of the form $\left\vert x,a\oplus f\left(
x\right)  \right\rangle $, where $\oplus$ denotes bitwise XOR and where I'm
ignoring workspace registers. \ However, if $f$ is injective, then another
model is possible: namely, an oracle that simply maps basis states of the form
$\left\vert x\right\rangle $ to basis states of the form $\left\vert f\left(
x\right)  \right\rangle $, \textquotedblleft erasing\textquotedblright\ the
previous contents of the $\left\vert x\right\rangle $ register. \ This second
model has the great advantage of not leaving $x$ around as garbage, but the
disadvantage of not being inherently reversible.

In 2000, Elham Kashefi (personal communication) asked me the following
question: are there sets of injective functions $f$ for which a quantum
computer can learn certain properties of $f$ using few queries to an erasing
oracle, but \textit{not} using few queries to a standard oracle? \ I realized
that a lower bound for the collision problem would naturally lead to an
affirmative answer to this question. \ This provided a central motivation for
my work on the collision problem \cite{aar:col}, which, in an appendix, did
give an affirmative answer to Kashefi's question.

More recently, I became aware that the converse question is equally interesting:

\begin{problem}
Are there sets of injective functions $f$ for which a quantum computer can
learn certain properties of $f$ using few queries to a standard oracle, but
\textit{not} using few queries to an erasing oracle?
\end{problem}

One natural candidate would be as follows:%
\[
f\left(  x\right)  =\left\langle h\left(  x\right)  ,\operatorname{gar}\left(
x\right)  \right\rangle ,
\]

\noindent where $h$ is a Simon function (that is, a function that's either
1-to-1 or else satisfies the Simon promise, and for which the promise is to
decide which), and $\operatorname{gar}\left(  x\right)  $ is a long string of
random garbage depending on $x$. \ The inclusion of $\operatorname{gar}\left(
x\right)  $ makes $f$ injective with overwhelming probability, but with a
standard oracle is no bar to running Simon's algorithm, since we can simply
use a second oracle invocation to uncompute garbage:%
\[
\left\vert x\right\rangle \rightarrow\left\vert x,h\left(  x\right)
,\operatorname{gar}\left(  x\right)  \right\rangle \rightarrow\left\vert
x,h\left(  x\right)  ,\operatorname{gar}\left(  x\right)  ,h\left(  x\right)
\right\rangle \rightarrow\left\vert x,h\left(  x\right)  \right\rangle .
\]

On the other hand, the garbage seems to make erasing queries no more useful
than classical queries. \ A central reason I'm interested in this conjecture
is that a proof of it seems likely to proceed by proving a much more general
statement, about the presence of a sufficient amount of garbage, in an erasing
oracle's responses, being equivalent (under appropriate conditions) to
decohering or measuring the responses.

\section{The Linear Cross-Entropy Benchmark}

In Fall 2019, a team at Google reported the achievement of quantum supremacy
based on a sampling benchmark with superconducting qubits \cite{arute}. \ In
Summer 2021, a team at USTC in China reported an independent replication
\cite{ustc}. \ Briefly, in these experiments, one generates a random quantum
circuit $C$ acting on $n$ qubits (in the Google experiment, $n=53$; in the
USTC experiment, $n=56$). \ One then uses a quantum computer to (hopefully)
sample from $\mathcal{D}_{C}$, the probability distribution over $n$-bit
strings induced by preparing the state $C\left\vert 0^{n}\right\rangle $ and
then measuring all $n$ qubits in the computational basis. \ Finally, having
generated samples $s_{1},\ldots,s_{k}\in\left\{  0,1\right\}  ^{n}$, one uses
a classical computer to calculate%
\[
\chi:=\frac{2^{n}}{k}\sum_{i=1}^{k}\left\vert \left\langle 0^{n}%
|C|s_{i}\right\rangle \right\vert ^{2}.
\]

One can show that ideal sampling, with a noiseless quantum computer, would
yield an expected value of $\chi\approx2$, whereas classical random guessing
would yield an expected value of $\chi\approx1$. \ The test is considered a
success if and only if $\chi$ is sufficiently bounded above $1$. \ Google's
experiment achieved a value of $\chi\approx1.002$.

One can ask: what if we wanted to achieve $\chi\gg2$? \ Would that problem be
intractable even for a quantum computer---analogous to violating the Tsirelson
inequality (i.e., the statement that even quantumly entangled players can win
the CHSH game with probability at most $\cos^{2}\frac{\pi}{8}$)? \ If we
imagined a black box able to output samples with, say, $\chi\approx3$, would
that black box provide \textquotedblleft beyond-quantum\textquotedblright%
\ computational abilities, and if so can we say anything about those abilities?

Here I ask a query complexity version of this question. \ Given a Boolean
function $f:\left\{  0,1\right\}  ^{n}\rightarrow\left\{  -1,1\right\}  $, an
easy quantum algorithm samples a string $s$ with probability equal to
$\widehat{f}\left(  s\right)  ^{2}$, where $\widehat{f}\left(  z\right)  $ is
the $z^{th}$ Boolean Fourier coefficient of $f$. \ Define the quantity%
\[
\chi:=\frac{2^{n}}{k}\sum_{i=1}^{k}\widehat{f}\left(  s_{i}\right)  ^{2}.
\]
Then one can show that repeating the Fourier sampling algorithm yields samples
$s_{1},\ldots,s_{k}$ that satisfy $\chi\approx3$. \ We now ask:

\begin{problem}
What is the quantum query complexity of outputting samples $s_{1},\ldots
,s_{k}$ that satisfy, say, $\chi\approx4$?
\end{problem}

Very recently, Kretschmer \cite{kretschmer} made significant progress on this
problem, by showing that

\begin{enumerate}
\item[(1)] given an $n$-qubit Haar-random \textit{quantum} oracle,
$\widetilde{\Omega}\left(  2^{n/4}\right)  $\ queries are needed to violate
the \textquotedblleft quantum supremacy Tsirelson's
inequality\textquotedblright\ for that oracle (compared to an upper bound of
$O\left(  2^{n/3}\right)  $), and

\item[(2)] when $k=1$, the obvious quantum algorithm for Fourier-sampling a
Boolean function $f$ is optimal among all $1$-query quantum algorithms.
\end{enumerate}

\section{Acknowledgments}

I thank Andrew Childs for the glued trees figure, and Travis Humble and
Mingsheng Ying for commissioning this piece.

\bibliographystyle{plain}
\bibliography{thesis}

\end{document}